\documentclass[12pt]{article}
\usepackage{amsxtra,amssymb,amsthm,amsmath,latexsym}

\theoremstyle{plain}

\newtheorem{lemma}{Lemma}[section]

\def\R{{\mathbb R}}

\def\C{{\mathbb C}}
\def\oH{\buildrel\circ\over H}
\def\oH1{\buildrel\circ\over H\kern-.02in{}^1}

\def\l{\ell}
\def\dotf{\dot{f}}

\def\ind{\hbox{ind}}

\def\Im{\hbox{\,Im\,}}

\begin{document}


\title{Reconstruction of the potential from $I$-function.
   \thanks{key words: inverse scattering, $I$-function, spectral function,
   scattering data }
   \thanks{Math subject classification: 34R30 }
}

\author{
A.G. Ramm\\
 Mathematics Department, Kansas State University, \\
 Manhattan, KS 66506-2602, USA\\
ramm@math.ksu.edu\\
}

\date{}

\maketitle\thispagestyle{empty}

\begin{abstract}
If $f(x,k)$ is the Jost solution and $f(k) = f(0,k)$, then the
$I$-function
is $I(k) := \frac{f^\prime(0,k)}{f(k)}$.
It is proved that $I(k)$ is in one-to-one
correspondence with the scattering triple
${\mathcal S} :=\left\{ S(k), k_j, s_j, \quad 1 \leq j \leq J\right\}$
and with the spectral function $\rho(\lambda)$ of the Sturm-Liouville
operator $l= -\frac{d^2}{dx^2} + q(x)$ on $(0, \infty)$ with the Dirichlet
condition at $x=0$ and
$q(x) \in L_{1,1} := \{q: q= \overline q, \int^\infty_0 (1+x) |q(x) dx <
 \infty\}$.

Analytical methods are given for finding $\mathcal S$ from $I(k)$ and
$I(k)$ from $\mathcal S$, and $\rho(\lambda)$ from $I(k)$ and
$I(k)$ from $\rho(\lambda)$. Since the methods for finding $q(x)$ from
$\mathcal S$ or from $\rho(\lambda)$ are known, this yields the methods for
finding $q(x)$ from $I(k)$.

\end{abstract}


\section{Introduction}
Let $q(x) \in L_{1,1} :=\{q : q= \overline q, \int^\infty_0 (1+x) |q(x)|
  dx < \infty\}$
and $\l u := - u^{\prime \prime} + q(x) u$ be the selfadjoint operator defined
in $L^2(0, \infty)$ by the boundary condition $u(0) = 0$. Let $\rho(\lambda)$
be its (uniquely defined) spectral function and
$${\mathcal S} := \left\{ S(k), k_j, s_j, 1 \leq j \leq J \right\}
        \eqno{(1.1)}$$
be its scattering data.

Here $k_j > 0, -k_j^2$ are the negative eigenvalues of $l$, $J$ is the
number of these eigenvalues, $s_j>0$ are the norming constants,
$$s_j := \| f_j (x) \|^{-2}, \quad \| \cdot \| = \| \cdot \|_{L^2(0,\infty)},
  \quad f_j(x) := f(x, ik_j),$$
$f(x,k)$ is the unique solution of the problem
$$(l - k^2) f(x,k) = 0, \quad x>0; \quad f(x,k) = e^{ikx} +  o(1) \hbox{\
as\ }
  x \to +\infty.$$
Let $f(k) := f(0,k)$. Then $S(k) := \frac{f(-k)}{f(k)}$.

It is known \cite{R1} that:
$$ d\rho (\lambda) = \begin{cases}
                        \frac{\sqrt{\lambda} d \lambda}
                        {\pi |f(\sqrt{\lambda})|^2}, \quad \lambda > 0 \\
                        \sum^J_{j=1} c_j \delta (\lambda + k^2_j)
                        d\lambda, \quad \lambda < 0,
                        \end{cases}
                        \eqno{(1.2)}$$
where $c_j := \| \varphi(x, ik_j) \|^{-2}$, the function $\varphi(x,k)$
is the unique solution to the problem:
$$(l - k^2) \varphi = 0, \quad x >0; \quad \varphi (0,k) = 0, \quad
   \varphi^\prime (0,k) = 1.
  \eqno{(1.3)}$$

The function $f(x,k)$ is an analytic function of $k$ in
$\C_+ := \{k : k \in \C, \Im k>0\}$, and $\varphi(x,k)$ is an entire
function of $k$. The numbers $ik_j, 1 \leq j \leq J$, are simple zeros of
$f(k)$ and $f(k)$ has no other zeros in $\C_+$, but it may have a simple
zero at $k=0$.

Define the I-function by the formula

$$I(k) := \frac{f^\prime (0,k)}{f(k)}. \eqno{(1.4)}$$
This function is of interest in applications and in mathematics: in
applications this function is an impedance function (a ratio of a component
of the electric field and a component of the magnetic field),
and in mathematics it is the Weyl function.

Recall that the Weyl function $m(k)$ is defined as such a function of
$k \in \C_+$ that
$$W(x,k) := \theta (x,k) + m(k) \varphi(x,k) \in L^2 (0, \infty), \quad
  \Im k > 0. \eqno{(1.5)}$$

Here $\varphi(x,k)$ was defined above and $\theta(x,k)$ is the solution
to the problem
$$(l - k^2) \theta = 0, \quad \theta (0,k) = 1, \quad \theta^\prime
  (0,k) = 0. \eqno{(1.6)}$$

If $q \in L_{1,1}$, then $W(x,k) = c(k) f(x,k), c(k) \not\equiv 0$.
Therefore $I(k) = \frac{W^\prime (0,k)}{W(k)} = m(k)$ as claimed.

The basic results of this paper are the methods and formulas for finding
${\mathcal S} (k)$ and $\rho(\lambda)$ from $I(k)$ and $I(k)$ from either
${\mathcal S}(k)$ or $\rho(\lambda)$.

Let us describe the results in more detail. Suppose ${\mathcal S}(k)$ is
known. Then, as we prove, $f(k)$ is uniquely and analytically determined by
solving the Riemann problem:
$$f(k) = S(-k) f(-k). \eqno{(1.7)}$$
This problem is solved in section 3, see formula (3.2).

If $f(k)$ is found, one finds $I(k),$ and 
therefore $f^\prime(0,k),$ by solving the
following problem:
$$\frac{f^\prime (0,k)}{f(k)} - \frac{f^\prime(0, -k)}{f(-k)} =
  \frac{2ik}{|f(k)|^2}, \eqno{(1.8)}$$
which is an immediate consequence of the Wronskian formula:
$$f^\prime (0,k) f(-k) - f^\prime(0,-k) f(k) = 2ik. \eqno{(1.9)}$$
Problem (1.8) is solved in section 3, see formulas (3.9) and (3.12).

If $\rho(\lambda)$ is given, then $I(k)$ is uniquely determined as follows:
one determines $|f(k)|$ and then analytically, $f(k)$, since $\rho(k)$
determines explicitly the numbers $k_j$ and $J$, where $1\leq j \leq J$.
The function $f(k)$ is determined analytically if its modulus on the real
axis and its zeros $ik_j, 1\leq j \leq J,$ are known.
 If $f(k)$ is found then $f'(0,k)$ can be found from the Riemann problem
(1.9). Alternatively, $I(k)$ can be
found as the solution of (1.8), which is a Riemann-type 
problem, the problem of finding a section-meromorphic function with
finitely many known simple poles, located at the points $ik_j, 1\leq j
\leq J,$ and known residues at these poles, from its
jump across the contour, the real axis in our case. In fact, one can solve 
(1.8) for $I(k)$ if the modulus of $f(k),$ the numbers $k_j,
1\leq j \leq J,$ and the residues (1.10) of the function $I(k)$ 
at its simple poles $ik_j$ are
known. 

Conversely, if $I(k)$ is known for all $k>0$, then $k_j$ are uniquely
determined since $k_j$ are the (simple) poles of $I(k)$ in $\C_+$. The number
$J$ of these poles is also uniquely defined. One has
$$I_j := \hbox{Res\ }_{k=ik_j} I(k) = \frac{f^\prime(0,ik_j)}{\dotf (ik_j)},
 \quad k_j > 0. \eqno{(1.10)}$$

It is known (see \cite{R1}, \cite{R2}) that
$$s_j = -\frac{2ik_j}{\dotf (ik_j) f^\prime (0, ik_j)}, \quad
  c_j = \frac{-2ik_j f^\prime (0, ik_j)}{\dotf (ik_j)}, \eqno{(1.11)}$$
so
$$s_j = -\frac{2ik_j}{\dotf^2(ik_j) I_j}, \quad c_j = -2ik_j I_j.
  \eqno{(1.12)}$$

To determine $\mathcal S$, or $\rho(\lambda)$, it remains to determine
$f(k)$ from $I(k)$.

This is done by solving a Riemann problem:
$$f(k) = \frac{k}{\Im I(k)}  \frac{1}{f(-k)}. \eqno{(1.13)}$$

In sections 2 and 3 some analytic formulas are derived
for the solutions of (1.8) and (1.13).

{\it Recovery of the potential $q(x)$ from $I(k)$ can be considered done,
if one
recovers either $\mathcal S$ or $\rho (\lambda)$ from $I(k)$.}

Indeed, if $\mathcal S$ is recovered, then the known procedure (see e.g.
\cite{R1}):
$${\mathcal S} \Rightarrow F \Rightarrow A \Rightarrow q \eqno{(1.14)}$$
recovers $q(x)$.

This procedure (the Marchenko method) is analyzed in \cite{R1} and {\cite{R2},
where it is proved that each step of this procedure is invertible
$$
{\mathcal S} \Leftrightarrow F \Leftrightarrow A \Leftrightarrow q.
$$
The function $F(x)$ is defined as
$$F(x) = \sum^J_{j=1} s_j e^{-k_jx} + \frac{1}{2\pi} \int^\infty_{-\infty}
        [1-S(k)] e^{ikx} dk. \eqno{(1.15)}$$

The kernel $A=A(x,y)$ is the transformation kernel:
$$f(x,k) = e^{ikx} + \int^\infty_x A(x,y) e^{iky} dy. \eqno{(1.16)}$$
It is related to $F(x)$ by the Marchenko equation:
$$A(x,y) + F(x+y) + \int^\infty_x A(x,s) F(s+y) ds = 0, \quad
  0 \leq x \leq y < \infty, \eqno{(1.17)}$$
which is uniquely solvable for $A(x,y)$ if $\mathcal S$ (and therefore $F(x)$)
comes from $q \in L_{1,1}$.

If $A(x,y)$ is found from (1.17), then
$$q(x) = -2 \frac{dA(x,x)}{dx}. \eqno{(1.18)}$$

If $\rho(\lambda)$ is found from $I(k)$, then the Gelfand-Levitan procedure
recovers $q(x)$:
$$\rho \Rightarrow L(x,y) \Rightarrow K(x,y) \Rightarrow q(x).$$
Here
$$L(x,y) = \int^\infty_{-\infty} \varphi_0 (x, \sqrt{\lambda}) \varphi_0
  (y, \sqrt{\lambda}) d [\rho(\lambda) - \rho_0 (\lambda)], \eqno{(1.19)}$$
$\varphi_0(x, \sqrt{\lambda}) =
 \frac{\sin(\sqrt{\lambda} x)}{\sqrt{\lambda}}, \rho_0 (\lambda)$
is the spectral function of the operator $\l_0$, that is, of operator
$\l$ with $q(x) = 0$, $d\rho_0 (\lambda) = 0$ for
$\lambda < 0, d \rho_0 (\lambda) = \frac{\sqrt{\lambda}d\lambda}{\pi},
 \lambda > 0$.

The kernel $K(x,y)$ is uniquely determined as the solution to the
Gelfand-Levitan equation:
$$K(x,y) + L(x,y) + \int^x_0 K(x,s) L(s,y) ds = 0, \quad 0 \leq y \leq x.
  \eqno{(1.20)}$$

If $K(x,y)$ is found from (1.20), then
$$q(x) = 2 \frac{dK(x, x)}{dx}. \eqno{(1.21)}$$
The kernel $K(x,y)$ is the transformation kernel:
$$\varphi(x, k) = \varphi_0 (x,k) + \int^x_0 K(x,y) \varphi_0 (y,k) dy, \quad
  k >0, \eqno{(1.21)}$$
where $\varphi(x,k)$ solves (1.3), and
$\varphi_0(x,k) = \frac{\sin(kx)}{k}$ solves (1.3) with $q(x) = 0$.

It is proved in \cite{R1}, \cite{R2} that
$$\rho(\lambda) \Leftrightarrow L(x, y) \Leftrightarrow K(x,y)
  \Leftrightarrow q, \eqno{(1.23)}$$
for a very wide class of $q(x)$, namely for the real-valued $q(x)$ for
which the equation $(l-z) u=0, \Im z>0$, has exactly one solution
$u \in L^2(0, \infty)$ (the limit point at infinity case).

In \cite{R3} a necessary and sufficient condition is obtained for a function
$I(k)$ to be the $I$-function corresponding to an operator $l$ with the
potential $q(x)$ which is locally $C^{(m)}$-smooth and a method is outlined
for the recovery of $q$ from $I(k)$.

In \cite{R2}, \cite{R4} various properties of $I(k)$ are studied.

The main results of this paper are:

1) it is proved that $I(k)$ is in one-to-one correspondence with the
scattering data $\mathcal S$ and with the spectral function $\rho(\lambda)$.

2) formulas are given for finding $I(k)$ from $\mathcal S$ or from
$\rho(\lambda)$ and vice versa
((2.11), (2.13), (2.15), (2.16), (3.2), (3.4), (3.5), (3.9), (3.12)).

\section{Recovery of $\mathcal S$ and $\rho$ from $I(k)$.}
The function $I(k)$ is meromorphic in $\C_+$. Its values on the real axis
determine uniquely $I(k)$ in $\C_+$.

Given $I(k) \forall k >0$, one finds its poles in $\C_+$. These poles
are exactly the points $ik_j, k_j>0$. $I(k)$ may have a pole at
$k=0$. If $k=0$ is a pole of $I(k)$, then $|I(0)|= \infty$ since
$f^\prime (0,0) \neq 0$. The number $J$ of the points $ik_j$ is uniquely
determined by $I(k)$.

Let us derive a formula for finding $f(k)$ from $I(k)$. Start with
formula (1.18), which we write as:
$$f(k) = \frac{k}{\Im I(k)} \frac{1}{f(-k)}. \eqno{(2.1)}$$

Define
$$w(k) := \prod^J_{j=1} \frac{k-ik_j}{k+ ik_j},\,\, \text { if }\,\, f(0)\neq 0,
\,\, w_1(k):=\frac k {k+i} w(k)\,\, \text { if }\,\, f(0)=0, 
\quad
  f_0 (k) := \frac{f(k)}{w(k)}. \eqno{(2.2)} $$

Note that $f_0 (k)$ is analytic in $\C_+$, has no zeros in $\C_+$, and
$f_0(\infty) = 1$. Also
$$\overline{w(k)} = w(-k) = w^{-1} (k), \hbox{\ and\ }
|w(k)| = 1 \hbox{\ if\ } k \in \R. \eqno{(2.3)}$$

Equation (2.1) can be written as
$$f_0 (k) = \frac{k}{\Im I(k)} \frac{1}{f_0(-k) w(k) w(-k)} =
  g(k) f_0^{-1} (-k), \eqno{(2.4)}$$
where
$$g(k) := \frac{k}{\Im I(k)} > 0, \quad k \in \R, \eqno{(2.5)}$$
and the relation $w(k)w(-k)=1$ for  $k\in \R$ was used.

The function $f_0^{-1} (-k)$ is analytic in
$\C_- := \{k: \in \C, \Im k < 0\}$,
does not have zeros in $\C_-,$ and
$\lim_{|k| \to \infty, k \in \C_-} f_0^{-1} (-k) = 1$.

Therefore $\ln f_0 (k)$ and $\ln f_0^{-1} (-k)$ are functions analytic in
$\C_+$ and $\C_-$, respectively, vanishing at infinity and
$$ \ln f_0 (k) = \ln g(k) + \ln f^{-1}_0 (-k). \eqno{(2.6)}$$
Therefore
$$\ln f_0 (z) = \frac{1}{2\pi i} \int^\infty_{-\infty}
  \frac{\ln g(s) ds}{s-z}, \quad \Im z>0, \eqno{(2.7)} $$
since (2.7) implies that the section-analytic function (2.7) satisfies
(2.6), and there is exactly one such function: if there were another one,
then their difference $\Phi (z)$ would be analytic in $\C_+$ and in
$\C_-$, $\Phi(\infty) = 0$ in $\C_+$ and in $\C_-$ and
$\Phi_+ (k) = \Phi_- (k)$, where
$\Phi_+ (k) = \Phi(k+ i0), \Phi_- (k) = \Phi (k - i0), k \in \R$.
Thus $\Phi(z)$ is analytic in $\C$ and $\Phi(\infty) =0$.

By the Liouville theorem, $\Phi (z) \equiv 0$. This argument is well-known
(see, e.g., \cite{G}), and is given to make the presentation self-contained.

Let us summarize the result:

\begin{lemma} 
Given $I(k)$, one finds $k_j$ and $J$, and then $f(z)$ in $\C_+$ :
$$f(z) = \exp \left( \frac{1}{2 \pi i} \int^\infty_{-\infty}
  \frac{\ln \frac{s}{\Im I (s)} ds}{s-z} \right) w(z), \quad
  \Im z >0, \eqno{(2.8)}$$
where $w(z)$ is defined in (2.2).
\end{lemma}
Note that $g(s) = g(-s)$, since $\Im I(s) = -\Im I(-s), s \in \R_-$,
so one has:
$$\int^\infty_{-\infty} \frac{\ln g(s)}{s-z} ds = 2 \int^\infty_0
  \frac{z \ln g(s) ds}{s^2 - z^2}, \quad \Im z > 0, \eqno{(2.9)}$$
and (2.8) can be written as:
$$f(z) = \exp \left(\frac{1}{i \pi} \int^\infty_0
  \frac{z \ln \frac{s}{\Im I(s)} ds}{s^2-z^2} \right) w(z), \quad
  \Im z> 0. \eqno{(2.10)}$$

To calculate $f(k)$, $k>0$, one takes $z = k + i0$ in (2.8) or (2.10)
and uses the well-known formula:
$$\frac{1}{s-k-i0} = P \frac{1}{s-k} + i \pi \delta (s-k), \eqno{(2.11)}$$
where $P\frac{1}{s}$ is the principle value of $\frac{1}{s}$ and $\delta(s)$
is the delta-function, to get from (2.8):
$$f(k) = \exp \left(\frac{1}{2 \pi i} \int^\infty_{-\infty}
  \frac{\ln \frac{s}{\Im I (s)} ds}{s-k} + \frac{1}{2} \ln
  \frac{k}{\Im I(k)} \right) w(k), k>0. \eqno{(2.12)}$$

Thus, given $I(k) \forall k >0$, we recovered uniquely
$f(k) \forall k>0, k_j$, and $J$. The $s_j$ are found by the first  formula
 (1.12) and (1.10). So $\mathcal S$ is recovered and
$q(x)$ is recovered by the procedure (1.14).

Formula (2.8) holds if $f(0)\neq 0$ and if $f(0)=0$.
If $f(0)=0,$ then one can suggest an equivalent to (2.8) formula which
looks differently. Namely, if $f(0)=0,$ then
 (2.4) holds with $g_1:=g\frac {k^2+1} {k^2}$ in place of
$g$ and $f_1(k):=\frac {f(k)}{w_1(k)}$ in place of $f_0(k).$
 Note that $0<|g_1(0)|<\infty$ if $f(0)=0$. Indeed, $\dot f(0)\neq 0$
and, by (2.1), $g(k)=|f(k)|^2,$ so $0<|g_1(0)|<\infty,$ as claimed. 
Formula (2.8) holds if $f(0)=0$ and if $\frac{s}{\Im I (s)} $ is replaced
by $\frac{s^2 +1}{s \Im I (s)}$ and $w(k)$ is replaced by $w_1(k)$.  

Let us show now how to recover $\rho(\lambda)$ from $I(k)$. As above, one
recovers $k_j$ and $J$. One has
$$|f(k)| = \left(\frac{k}{\Im I(k)} \right)^{\frac{1}{2}}, \quad k>0,
  \eqno{(2.13)}$$
where the square root is positive and the function
$\frac{k}{\Im I(k)} >0, k>0$.

To find $c_j$ (see formula (1.2)) one uses second formula (1.12)
and formula (1.10). If
$\rho(\lambda)$ is found, then $q(x)$ is found by the Gelfand-Levitan
method (1.23).

\section{Recovery of $I(k)$ from $\mathcal S$ or $\rho$.}
{\it Let us first explain how to calculate $I(k)$ given $\mathcal
S$.}

 Assume
$\mathcal S$ is given. Then $f(k)$ can be recovered analytically as
follows.

Write (1.7), using (2.3), as
$$f_0 (k) = S(-k) w^{-2} (k) f_0 (-k),  \,\, f(0)\neq 0. \eqno{(3.1)}$$
The function $f_0(k)$ is analytic and has no zeros in $\C_+$, and
$f_0(\infty) =1$, and $f_0(-k)$ has similar properties in $\C_-$. Therefore
one solves analytically the Riemann problem (3.1) and gets:
$$f(z) = \exp \left(\frac{1}{2 \pi i} \int^\infty_{-\infty}
  \frac{\ln \frac{S(-t)}{w^2(t)} dt}{t-z} \right) w(z), \Im z>0.
  \eqno{(3.2)}$$

Note that the index $\ind_\R \frac{S(-k)}{w^2 (k)} = 0$
if $f(0)\neq 0$, so $f(z)$
is uniquely defined by formula  (3.2). As in section 2, one calculates
$f(k)$ for $k>0$ by taking $z = k+i0$ in (3.2) and using (2.11). 

Formula (3.2) holds also when $f(0)=0$. In this case 
$\ind_\R \frac{S(-k)}{w^2 (k)} = 1,$ and 
$\ind_\R S_1(-k) =0,$ where $S_1(k):=S(k)\frac {k-i}{k+i}.$
To see that formula (3.2) holds in the case $f(0)=0,$ one can use
the above argument and the following formula:
$\frac 1 {2\pi i}\int_{-\infty}^{\infty}\frac {\log \frac
{s-it}{s+it}}{s-z}ds=log\frac z {z+it},$ which holds if $t>0$
and $\Im z>0.$

Alternatively, if $f(0)=0$, then one replaces in formula (3.1) $f_0(k)$ 
by $f_1:=\frac{f(k)}{w_1(k)},$ where $w_1$ is defined in (2.2), and $S(k)$ by
$S_1(k):=S(k)\frac {k-i}{k+i}.$ Formula (3.2) in this case takes the
form:
$$f(z) = \exp \left(\frac{1}{2 \pi i} \int^\infty_{-\infty}
  \frac{\ln \frac{S_1(-t)}{w^2(t)} dt}{t-z} \right) w_1(z), \Im z>0,
  \eqno{(3.2')}$$ 
equivalent to (3.2).

Since $f(k)$ is found, to recover $I(k)$ one can use formula (1.8):
$$I(k) = I(-k) + \frac{2ik}{|f(k)|^2}\,. \eqno{(3.3)}$$
The function $I(k)$ is meromorphic in $\C_+$. Let us subtract from
$I(k)$ its principal parts at the poles and get a holomorphic 
in $\C_+$ function ${\mathcal J}(k)$ defined as: 
$${\mathcal J}(k) := I(k) - ik - \sum^J_{j=0} \frac{I_j}{k-ik_j},
  \eqno{(3.4)}$$
where $k_0 := 0$ and the term with $j=0$ is included only if $f(0) = 0$.
The numbers $I_j, j>0$ are calculated by the first formula (1.12) if
$\mathcal S$ and $f(k)$ are known. The number
$I_0 = \frac{f^\prime(0,0)}{\dotf (0)}$. This number can be calculated if
one knows how to calculate $f^\prime (0,0)$.

One has, differentiating (1.16):
$$f^\prime (0,k) = ik - A(0,0) + \int^\infty_0 A_x (0,y) e^{iky} dy.$$
The number $A(0,0)$ can be calculated:
$$\begin{aligned}
  f(k) = 1+ \int^\infty_0 A(0,y) e^{iky} dy = 1 + A(0,y)
  \frac{e^{iky}}{ik} |^\infty_0 \\
  -\frac{1}{ik} \int^\infty_0
  A_y (0,y) e^{iky} dy = 1 - \frac{A(0,0)}{ik} - \frac{1}{ik}
  \int^\infty_0 A_y (0,y) e^{iky} dy.
  \end{aligned}$$
Thus
$$A(0,0) = - \lim_{k \to \infty} [ik (f(k) - 1)] \eqno{(3.5)}$$
To calculate $f^\prime (0,0)$, we divide the Wronskian formula (1.9) by
$k$ and let $k\to 0$. This yields:
$$f^\prime (0,0) = \frac{-i}{\dotf (0)},
  \eqno{(3.6)}$$
so that the residue of $I(k)$ at $k=0$, when $f(0) = 0$, is:
$$I_0 = \frac{f^\prime (0,0)}{\dotf (0)} = -\frac{i}{[\dotf (0)]^2}.
  \eqno{(3.7)}$$
Existence of $\dotf (0)$ for $q \in L_{1,1}$ is proved in \cite{R1}.

The function ${\mathcal J}(k)$ solves the following Riemann problem:
$${\mathcal J}(k) = {\mathcal J}(-k) + \frac{2ik}{|f(k)|^2} - 2ik -
  \sum^J_{j=0} I_j
  \left(\frac{1}{k-ik_j} + \frac{1}{k+ik_j} \right), k \in \R,
  \eqno{(3.8)}$$
and ${\mathcal J}(k)$ is analytic in $\C_+$, ${\mathcal J}(\infty) = 0$,
while ${\mathcal J}(-k)$ has similar
properties in $\C_-$. Thus (3.8) implies
$${\mathcal J}(z) = \frac{1}{\pi} v.p. \int^\infty_{-\infty}
\frac{dt}{t-z}
  \left[t\left(|f(t)|^{-2} - 1 \right) +i \sum^J_{j=0} I_j t
(t^2+k_j^2)^{-1}
  \right], \quad \Im z  > 0, \eqno{(3.9)}$$
and taking $z=k+ i0, k>0$, one can calculate ${\mathcal J}(k)$ for $k>0$
using formula (2.11).

If ${\mathcal J}(k)$ is found for all $k>0$, then $I(k)$ is calculated by
formula (3.4).

{\it Therefore $\mathcal S$ determines $I(k)$ uniquely
and analytically, since $\mathcal S$ determines $k_j$, $J,$ $I_j,$
and ${\mathcal J}(k)$ uniquely.}

{\it Let us explain how to calculate $I(k)$ given $\rho(\lambda)$.}

If $\rho(\lambda)$ is given, then (see formula (1.2)), the function
$|f(k)|$ is known for all $k>0$, and the numbers $k>0$, and the numbers
$k_j$, $J$ and $c_j$ are known.

Therefore the function $f(z)$ can be calculated analytically. Indeed,
$|f_0 (k)| = |f(k)|$ if $k \in \R$, where
$f_0(k) = \frac{f(k)}{w(k)}$. The function $f_0(z)$ is analytic and
has no zeros in $\C_+$, and $f_0(\infty) = 1$. Therefore
$\ln f_0 (z)$ is analytic in $\C_+$ and vanishes at infinity. So it can be
recovered by the Schwarz formula:
$$\ln f_0 (z) = \frac{1}{i \pi} \int^\infty_{-\infty}
  \frac{\ln |f_0 (t)|}{t-z} dt, \quad \Im z>0, \eqno{(3.10)}$$
which constructs an analytic function
in $\C_+$  given its real part on the
real axis. Since $|f_0 (t)|= |f(t)|$, one gets
$$f(z) = w(z) \exp \left\{ \frac{1}{i \pi} \int^\infty_{-\infty}
  \frac{\ln |f(t)|dt}{t-z} \right\}, \quad \Im z > 0. \eqno{(3.11)}$$

Taking $z=k+i0, k>0,$ in (3.11), and using (2.14) yields
$f(k)$ for $k>0$. One calculates the numbers $I_j$ by the second formula
(1.12).

If $k_j, J, I_j$ and $f(k)$ are known, then one calculates
${\mathcal J}(z)$ by formula (3.9), and ${\mathcal J}(k)$ for 
all $k>0$ by taking
$z = k+ i0$ in (3.9) and using (2.11). If 
${\mathcal J}(k)$ is known then $I(k)$ is
calculated by formula (3.4).

{\it This completes the description of the formulas for finding $I(k)$
given $\mathcal S$ or $\rho(\lambda)$.}

In conclusion let us make a remark concerning {\it numerical aspects} of
finding
$k_j$ given $I(k)$.

Since $I(k) = \overline{I(-k)}$ for $k \in \R$, the knowledge of
$I(k)$ for $k>0$ yields the values of $I(k)$ for all $k \in \R$. The function
$I(k)$ is meromorphic in $\C_+$ and is of the form (see formula (3.4)):
$$ I(k) = ik + \sum^J_{j=0} \frac{I_j}{k-ik_j} + {\mathcal J}(k),
  \eqno{(3.12)}$$
where ${\mathcal J}(k)$ is analytic in $\C_+$ and is
$o\left(\frac{1}{|k|} \right)$ as $|k| \to \infty$, $k \in \C_+$.
Therefore the numbers $J$, $I_j$ and $ik_j$ can be calculated 
if $I(k)$ is known for all $k>0$.
A method for calculating $J$, $I_j$ and $k_j$ can be based on
the formula:
$$\frac{1}{2 \pi i} \int^\infty_{-\infty} e^{ikt} \left[ I(k) - ik \right]
  dk =\frac {I_0}{2}+ \sum^J_{j=1} I_j e^{-k_jt}, \quad t >0. \eqno{(3.13)}$$

The behavior of $I(k)$ as $k \to +\infty$ one can obtain from the formulas:
$$f^\prime (0,k) = ik - A(0,0) + \widetilde A_1 (k), \quad
  \widetilde A_1 (k) :=
  \int^\infty_0 A_x (0,y) e^{iky} dy$$
$$f(k) = 1 + \widetilde A (k), \quad  \widetilde A(k) = \int^\infty_0 A(0,y)
  e^{iky} dy.$$
The functions $A(y) := A(0,y)$ and $A_1(y) := A_x (0,y)$ belong to
$L^1(0, \infty)$. Thus
$$I(k) = \frac{f^\prime (0,k)}{f(k)} = [ik - A(0,0) + \widetilde A_1]
  [1 + \widetilde (k)]^{-1} = ik -A (0,0) - ik \widetilde A + \dots =ik
  + o(1) \eqno{(3.14)}$$
where the dots stand for the terms of higher order of smallness as
$k \to + \infty$. The important point is: the constant term $-A(0,0)$ is
cancelled in the asymptotics since
$$-ik\widetilde A = A(0,0) + o(1), \hbox{\ as\ } k \to +\infty,
  \eqno{(3.15)}$$
as follows from the calculation preceding formula (3.5).

\end{document}